\documentclass[journal]{IEEEtran}
\usepackage{graphicx}
\usepackage{lipsum}  
\usepackage{tabularx}
\usepackage{adjustbox}
\graphicspath{ {./images/} }
\usepackage{comment}
\usepackage{url}
\usepackage{multirow}
\usepackage{multicol}
\usepackage{booktabs}
\usepackage{amsmath}
\usepackage{chngpage}
\usepackage{threeparttable}
\usepackage{bbm}
\usepackage{color}
\usepackage[hidelinks]{hyperref}
\usepackage{tikz}
\usepackage{calc}
\def\checkmark{\tikz\fill[scale=0.4](0,.35) -- (.25,0) -- (1,.7) -- (.25,.15) -- cycle;} 

\begin{document}
\title{Empowering cyberphysical systems of systems with intelligence}

\author{
\IEEEauthorblockN{Stavros Nousias$^{1}$, Nikos Piperigkos$^{1}$, Gerasimos Arvanitis$^{1,2}$, Apostolos Fournaris,  Aris S. Lalos$^{1}$ and Konstantinos Moustakas$^{2}$}\\
\IEEEauthorblockA{$^1$Industrial Systems Institute, Athena Research Center, Patras Science Park, Greece\\
$^2$Dept. of ELectrical and Computer Engineering, University of Patras, Greece
}
}


\markboth{Journal of \LaTeX\ Class Files,~Vol.~14, No.~8, August~2015}%
{Shell \MakeLowercase{\textit{et al.}}: Bare Demo of IEEEtran.cls for IEEE Journals}

\maketitle

\begin{abstract}
Cyber Physical Systems have been going into a transition phase from individual systems to a collecttives of systems that collaborate in order to achieve a highly complex cause, realizing a system of systems approach. The automotive domain has been making a transition to the system of system approach aiming to provide a series of emergent functionality like traffic management, collaborative car fleet management or large-scale automotive adaptation to physical environment thus providing significant environmental benefits (e.g air pollution reduction) and achieving significant societal impact. Similarly, large infrastructure domains,  are evolving into global, highly integrated cyber-physical systems of systems covering all parts of the value chain. In practice, there are significant challenges in CPSoS applicability and usability to be addressed, i.e. even a small CPSoS such as a car consists several subsystems Decentralization of CPSoS appoints tasks to individual CPSs within the System of Systems. CPSoSs are heterogenous systems. They comprise of various, autonomous, CPSs, each one of them having unique performance capabilities, criticality level, priorities and pursued goals. all CPSs must also harmonically pursue system-based achievements and collaborate in order to make system-of-system based decisions and implement the CPSoS functionality. This survey will provide a comprehensive review on current best practices in connected cyberphysical systems. The basis of our investigation is a dual layer architecture encompassing a perception layer and a behavioral layer. Perception algorithms with respect to scene understanding (object detection and tracking, pose estimation), localization mapping and path planning are thoroughly investigated. Behavioural part focuses on decision making and human in the loop control.

\end{abstract}


%
\IEEEpeerreviewmaketitle

\section{Introduction}
The Cyber-Physical System domain the past few years have been going into a transition phase from individual systems operating isolated to a collection of systems that collaborate to achieve a highly complex cause, realizing a system of systems approach. There is a significant investment in Cyber-Physical Systems of Systems both within or outside Europe for domains, like automotive, industrial manufacturing, railways, aerospace, smart buildings, logistics, energy, industrial processes, that have a significant impact on the European economy and society. 

The automotive domain, for example, that provides jobs for 12 million people and accounts for 4\% of the EU’s GDP including sales and maintenance for 4.3 million, and transport for 4.8 million, has been thoroughly investing in CPSs inside cars either to provide elaborate control for traditional automotive processes (like brake system, steering etc) or to introduce new concepts like autonomous/semi-autonomous driving. This domain has been making a transition to the system of system approach aiming to provide a series of emergent functionality like traffic management, collaborative car fleet management or large-scale automotive adaptation to the physical environment thus providing significant environmental benefits (e.g air pollution reduction) and achieving significant societal impact.

Similarly, large infrastructure domains, like industrial manufacturing with more than 30 million employees, a turnover of € 6,410 billion, and a value-added of € 1,590 billion in 2010 are evolving into global, highly integrated cyber-physical systems of systems that go beyond pure production and that covers all parts of the value chain, including research, design, and service provision. This novel approach can enable a high level of flexibility that can be interpreted to fast adaptation to customer requirements, a high degree of product customization and better industrial sustainability. 


Achieving collective behaviour for CPSoS based solutions in large scale control processes will help citizens improve their quality of life through smart, safe, and secure cities, energy-efficient buildings and cars, and green infrastructures (traffic management, lighting, water and waste management); and smart devices and services for smart home functionality, home monitoring, health services, and assisted living.

However, in practice, there are significant challenges in CPSoS applicability and usability to be addressed to take full advantage of the CPSoS benefits and sustain/extend their growth. The fact that even a small CPSoS, (eg. a connected car) consists of several subsystems and executes thousands of lines of code highlights the complexity of the system of system solution and the extremely elaborate CPSoS orchestration which highlights the need for an approach beyond traditional control and management center\cite{engell2015core}. 

Given this, having a centralized authority that handles all CPSoS processes, subsystems and control loops seems to be very hard to capture and implement thus pointing to a different design, control and management approach. Decentralization of CPSoS processes and overall functionality by appointing tasks to individual CPSs within the System of Systems can be a reasonable solution, yet still, the collaborative mechanism between CPSs (that constitute the CPSoS behaviour) remains a point of research since appropriate tools and methodologies are needed to assess that the expected system of system functional requirements is retained (the CPSoS operates as it should be) and the non-functional requirements are matched (the CPSoS remains resilient, safe and efficient).

CPSoSs are heterogeneous systems. They consist of various, autonomous, CPSs, each one of them having unique performance capabilities, criticality level, priorities and pursued goals. CPSs in general are self-organized and, on several occasions, they may have conflicting goals thus competing to get access to common resources. However, from a CPSoS perspective, all CPSs must also harmonically pursue system-based achievements and collaborate to make system-of-system based decisions and implement the CPSoS behaviour. Considering that CPSoS consists of many CPSs, finding the methodology to achieve such an equilibrium in a decentralized way is not an easy task. The above issue becomes more complex when we also consider the amount of data to be exchanged between CPSs and the processing of those data. The collection of data and the data analytics need to be refined in such a way that only the important information is extracted and forwarded to other CPSs and the overall system. Also, Mechanisms to handle, in a distributed way, large amounts of data are needed to extract cognitive patterns and detect abnormalities. Thus, some data classification, labelling and refinement mechanisms should be put in place locally (in each CPS) to offload the complexity and communication overhead at the system of system level\cite{atat2018big}.

In the above-described setup, we cannot overlook the fact that CPSoS depend on humans since humans are part of the CPSoS functionality and services, interact with the CPSs and contribute to the CPSoS behaviour. Operators and managers play a key role in the operation of CPSoS and take many important decisions while in several cases human CPS users are the key player in the CPSoS main role (thus forming Cyber-Physical Human Systems). Thus, we need to structure a close symbiosis between computer-based systems and human operators/users and constantly enhance human situational awareness as well as devise a collaborative mechanism on handling CPSoS decisions, forcing the CPSoS to comply with human guidelines and reactions. Novel approaches on Human Machine Interfaces that employ eXtended Reality (XR) principles need to be devised to help humans gain fast and easy to grasp insight into the CPSoS processes but also to enrol them seamlessly to the CPSoS operation.

Finally, it cannot be overlooked that security and trust in CPSoS operation must be retained at all costs since CPSoSs are physically entangled systems and are in close integration with humans. Security breaches can lead to serious incidents that may affect human lives (in automotive, energy, aerospace, railways, industrial domain etc.). The autonomous nature of the CPSoS, the high heterogeneity and the use of legacy components, however, makes traditional security measures hard to apply, thus highlighting the need for a new, CPS applicable security and trust mechanism that must be applied to the system from design time and follow the CPSoS design operate continuum, thus constantly be updated, reconfigured and redesigned according to cybersecurity abnormalities. To achieve that, security components must be modelled based on the security-by-design principle considering that they may be placed in CPSs with various, different, security needs and performance capabilities. Such components must be realized during the design/redesign of the CPSoS while in parallel specialized security monitoring mechanisms and tools must be introduced in the autonomous CPSs and the System as a whole, so that they can detect, identify, respond and mitigate a security attack in the presence of unforeseen conditions that may emerge during CPSoS operation including resilience failures. 

This survey will provide a comprehensive review of current best practices in connected cyber-physical systems. The basis of our investigation is a dual-layer architecture encompassing a perception layer and a behavioural layer. Perception algorithms concerning scene understanding (object detection and tracking, pose estimation), localization mapping and path planning are thoroughly investigated.
The behavioural part focuses on decision making and human in the loop control. 
The rest of this survey is organized as follows. Section II analyses aspects of the perception layer focusing on 2D and 3D object detection and scene analysis, localization, SLAM and path planning and human-centric perception.
Section III focuses on the behavioural layer, while Section IV concludes this survey paper.


\section{Perception layer}
\subsection{Object detection and scene analysis }
\subsubsection{Object detection from 2D images}
Object detection has been evolved considerably since the appearance of deep convolutional neural networks \cite{zhao2019object}.
Nowadays, there are two main branches of proposed techniques. In the first one, the object detectors, using two stages, 
generate region proposals
which are subsequently classified in the categories that are determined by the application at hand (e.g., vehicles, cyclists and
pedestrians, in the case of autonomous driving). Some important, representative, high-performance examples of this first branch are 
Faster R-CNN \cite{ren2016faster}, Region-based Fully Convolutional Network (R-FCN) \cite{dai2016r}, Feature Pyramid Network (FPN) \cite{lin2017feature} and Mask R-CNN \cite{he2017mask}. In the second branch, object detection is cast to a single-stage, regression-like
task with the aim to provide directly both the locations and the categories of the detected objects. Notable examples, here,
are Single Shot MultiBox Detector (SSD) \cite{liu2016ssd}, SqueezeDet \cite{wu2017squeezedet}, YOLOv3 \cite{redmon2018yolov3} and EfficientDet \cite{tan2020efficientdet}.

\color{black}

\subsubsection{Object detection from 3D images}
Object detection in LIDAR point clouds is a three-dimensional problem where the sampled points are not uniformly distributed over the objects in the scene and do not directly correspond to a cartesian grid. 
3D object detection is dominantly performed with 3D convolutional networks due to the irregularity and lack of apparent structure in the point cloud.
Several transformations take place to match the point cloud to feature maps that are forwarded into deep networks. Commendable detection outcomes appear in the literature as early as 2016. Li et al.\cite{li2016vehicle} projected the 3D points in a 2D map and employed 2D fully convolutional networks to successfully detect cars in a LIDAR point cloud reaching an accuracy of 71.0\% for car detection of moderate difficulty. A follow-up paper \cite{li20173d} proposes 3D fully convolutional networks reporting accuracy of 75.3\% for car detection of moderate difficulty. However, since dense 3D fully convolutional networks demonstrate high execution times, Yan et al.\cite{yan2018second} investigated an improved sparse convolution method for such networks, which significantly increases the speed of both training and inference. According to KITTI benchmarks, the reported accuracy reaches 78.6\% for car detection of moderate difficulty. To revisit 2D convolutions in 3D object detection Pointpillars \cite{lang2019pointpillars} proposed a novel encoder that utilizes PointNets to learn a representation of point clouds organized in vertical columns (pillars) and subsequently employed a series of 2D convolutions. Pointpillars reported accuracy of 77.28\% in the same category. Shi et al. proposed PointRCNN \cite{shi2019pointrcnn} for 3D object detection from raw point cloud. They devised a two stages stage approach where the first stage yields a bottom-up 3D proposal generation the second stage refines the proposals in the canonical coordinates to obtain the final detection results, reporting accuracy of 78.70\%. An extended variation of PointRCNN is the part-aware and aggregation neural network(Part-$A^2$ Net). The part-aware stage for the first time fully utilizes free-of-charge part supervisions derived from 3D ground-truth boxes to simultaneously predict high-quality 3D proposals and accurate intra-object part locations. Then the part-aggregation stage learns to re-score the box and refines the box location by exploring the spatial relationship of the pooled intra-object part locations. The reported accuracy reaches 79.40\%.

\color{black}
\subsection{Localization, SLAM and Path planning}
Unmanned vehicles, either ground (UGV), aerial (UAV) or underwater (UUV), are prominent CPSoS. Typical examples include autonomous vehicles and robots, operating for a variety of different civilian and military challenging tasks. At the same time, the prototyping of 5G and V2X (e.g. V2V and V2I) related communication protocols enable the close collaboration of vehicles, to address their main individual or collective goals. Autonomous vehicles with inter-communication and network abilities are known as Connected and Automated Vehicles (CAV), being part of the more general concept of Connected CPSoS. The main focus of CAV's related technologies is to increase and improve safety, security and energy consumption of (cooperative or not) autonomous driving, by the strict control of vehicle's position and motion \cite{Montanaro2018}. At a higher level, CAV have the potential for a further enhancement of the transportation sector's overall performance. 

Perception and scene analysis ability are fundamental for a vehicle's reliable operation. Computer vision-based object detection and tracking should be seen as a first (though necessary) pre-processing step, feeding more sophisticated operational modules of vehicle \cite{Eskandarian2021}. The latter is imperative to have accurate knowledge of both its own and its neighbours' (vehicles, pedestrians or static landmarks) position, in order to design efficiently the future motion actions, i.e. to determine the best possible velocity, acceleration, yaw rate, etc. These motion actions primarily focus on e.g. keeping safe inter-vehicular distances, eco-friendly driving by reducing gas emissions, etc. The above challenges can be addressed in the context of Localization, SLAM and Path planning, which are discussed below:
\subsubsection{Localization} Localization module is responsible for providing absolute position information to the vehicles. Global Navigation Satellite Systems (GNSS), like GPS, Beidou, Glonass, etc., are usually exploited for that purpose. GPS sensor is currently employed as the most common commercial device. It is straightforward to couple or fuse GPS information with Inertial Measurement Unit (IMU) readings \cite{Noureldin2013}, to design a complete Inertial Navigation System (INS) providing positioning, velocity and timing solutions (PVT). IMU sensor consists of gyroscopes and accelerometers for measuring yaw rate and acceleration (in $x$,$y$,$z$ directions) of vehicle. Additionally, odometers and wheel sensors \cite{Skog2009} can also be utilized. However, even highly reliable IMU sensors suffer from accumulative or drift error, significantly reducing their consistency as the vehicle is moving. Another limitation of stand-alone GPS localization is directly related to GPS itself. Its accuracy is highly degraded in dense urban canyons or tunnels \cite{Kuutti2018},  even exceeding 10$m$ error. Main sources of GPS signal degradation are due to \cite{Noureldin2013} satellite clock error, receiver clock error, ionosphere delay, tropospheric delay, multi-path, etc. Moreover, it is vulnerable to cyber-attacks \cite{Ren2020}, like spoofing or jamming. The former causes an intentionally "wrong" position, even kilometres away from expected GPS measurement. The latter poses a rather more severe threat since it totally blocks the GPS signal. Several alternative approaches relying on ground base stations have been developed for enhancing localization accuracies, such as Assisted GPS (AGPS) or Differential GPS (DGPS). However, they are also susceptible to multi-path effect and signal blockage \cite{Alam2013}. The desired localization error, as it has been reported in the literature, should be lower than  1$m$ (where-in-lane accuracy) \cite{Neto2020} to meet the standards of autonomous driving. For example, if a vehicle is localized on the curb instead of the road, it may lead to a serious accident with pedestrians or other vehicles. Therefore, it is quite clear that for obtaining the desired positioning solutions, other types of advanced sensors, like LIDAR, Camera, RADAR, etc., must be additionally taken into account. Moreover, the emergence of V2V communications in the context of the Internet of Things (IoT), facilitates the exploitation of both onboard and off-board information, in order to design a more robust localization system. This collaborating multi-modal fusion of heterogeneous measurements is known as Cooperative Localization (CL), a rather recent and very promising technique that is able to tackle the limitations and drawbacks of GPS/IMU localization. Each vehicle can now receive external information (like absolute position, relative distance, velocity, acceleration, etc.) from nearby vehicles, infrastructure or pedestrians, effectively assisting its own localization system.
    
    There are many existing works \cite{Kuutti2018}, \cite{Buehrer2018}, \cite{Wymeersch2009}, \cite{Safavi2018}, \cite{Gao2019} that survey related aspects, challenges and algorithms of CL. For example, \cite{Kuutti2018} provides an overview of current trends and future applications of Localization (not only CL) in autonomous vehicles environments. The discussed techniques are mainly distinguished on the basis of the utilized sensor. Ranging measurements like relative distance and angle can also be extracted through the V2V abilities of CL. Common ranging techniques include Time of Arrival (TOA), Angle of Arrival (AoA), Time Difference of Arrival (TDOA), Received Signal Strength (RSS), etc. The works of \cite{Buehrer2018}, \cite{Wymeersch2009} delve into detailed mathematical modelling of CL tasks. More specifically, \cite{Buehrer2018} exploits various criteria to categorize related algorithms: 
    \begin{enumerate}
        \item Measurement type: The sensor or ranging technique being used for localization. V2V communications enable different ranging methods to be used (as mentioned above).
        \item Centralized vs Distributed: Centralized algorithms require nodes/vehicles of the network to broadcast their measurements to a fusion centre (e.g. cloud or some leader-vehicle), responsible for all the computations. Although higher accuracy can be achieved, limitations like communication overhead, computational power, network size, fusion centre malfunctioning etc, must be taken into account. On the contrary, with distributed processing architecture, the computations are assigned to each vehicle which interacts only with close neighbours.  
        \item One-shot vs Tracking: One-shot refer to methods that do not exploit any past information. On the other hand, tracking has to do with algorithms that, apart from measurements, employ kinematic models in order to approximate the actual movement of vehicles. Tracking methods exploit Bayesian estimators as mentioned below.
        \item Fusion estimator: Multi-modal fusion is vital for increased location estimation accuracy. Fusion can be effectively performed using well-known estimators like Least Squares (LS), Maximum Likelihood (ML), Minimum Mean Square Error (MMSE), Maximum A Posteriori (MAP), etc. One-shot ML estimator coincides with (weighted by measurement noise variance) LS when the measurements are corrupted by Gaussian noise. MMSE and MAP are common Bayesian estimators which treat the unknown vehicle's position as a random variable, instead of a deterministic value as one-shot do. Kalman, Extended Kalman and Unscented Kalman Filters (KF, EKF, UKF) are prominent examples of MMSE estimators. Belief propagation and factor graphs optimization is also an important MAP tool.
    \end{enumerate}
    The \cite{Wymeersch2009} formulates a distributed Gradient Descent (GD) algorithm as LS solution and the Bayesian factor graph approach of Sum Product Algorithm over Wireless Networks (SPAWN). In general, distributed and tracking/Bayesian algorithms are more attractive to perform CL. An overview of distributed localization algorithms in IoT is also given in \cite{Safavi2018}. The authors discuss in addition the proposed distributed geometric framework of DILOC, as well as the extended versions of DLRE and DILAND, which facilitates the design of a linear localization algorithm. These methods require the vehicle to be inside the convex full formed by 3 neighbouring anchors (nodes with known and fully accurate positions) and to compute its barycentric coordinates with respect to neighbours. However, major challenges are related to mobile scenarios due to varying topologies, as well as how feasible the presence of anchors will be in automotive applications. An interesting approach is discussed in \cite{Meyer2016}, where mobile agents in general, try to cooperatively estimate their position as well as to track non-cooperative objects. The authors developed a distributed particle filter-based belief propagation approach with message passing, though they consider the presence of anchor nodes. Furthermore, the computational and communication overhead may be a serious limitation towards real-time implementation. In \cite{Soatti2018}, a novel distributed technique is proposed to improve the stand-alone GNSS accuracy of vehicles. Once again, noncooperative objects or features (e.g. trees, pedestrians, etc.) are exploited in order to improve location accuracy. Features are cooperatively detected by vehicles using their onboard sensors (e.g. LIDAR), where a perfect association is assumed. These Vehicle-to-Features measurements are fused with GNSS in the context of a Bayesian message-passing approach and KF. Experimental evaluation was assessed using SUMO simulator, however, the number of detected features, as well as communication overhead, should be taken into serious account. The work of \cite{Brambilla2020} extends \cite{Soatti2018}, by proposing a distributed data association framework for features and vehicles. Data association was based on belief propagation. Validation was performed in realistic urban traffic conditions. The main aspect of \cite{Brambilla2020}, as well as \cite{Soatti2018}, is that vehicles must reach a consensus about features state, in order to improve their location. Graph Laplacian CL has been introduced in \cite{Piperigkos2020}, \cite{Piperigkos2020b}. Centralized or distributed Laplacian Localization formulates a LS optimization problem, which fuses the heterogeneous inter-vehicular measurements along with the V2V connectivity topology through the linear Laplacian operator. EKF and KF based solutions have been proposed for addressing CL in tunnels \cite{Elazab2017}, \cite{Yang2020} when the GPS signal may be blocked. A distributed robust cubature KF enhanced by Huber M-estimation is presented in \cite{Liu2017}. The method is used to tackle the challenges of data fusion under the presence of outliers. Pseudo-range measurements from satellites are also considered during the fusion process. Authors in \cite{Zhao2020} developed a distributed Bayesian CL method for localizing vehicles in the presence of Non-Line-of-Sight range measurements and spoofed vehicles. They focused primarily on ego vehicle location estimation and abnormal vehicles detection rates.
    \subsubsection{SLAM}
    Simultaneous Localization and Mapping (SLAM) is also a relevant task of Localization. It refers to the problem of mapping an environment using measurements from sensors (e.g. Camera or LIDAR) on-board the vehicle or robot, whilst at the same time estimating the position of that sensor relative to the map. Although when stated in this way SLAM can appear to be quite an abstract problem robust and efficient solutions to the SLAM problem are critical to enabling the next wave of intelligent mobile devices. SLAM in its general form tries to estimate over a time period, the poses of the vehicle/sensor and the landmarks’ position of the map, given control input measurements, provided by odometry sensors onboard the vehicle and measurements with respect to landmarks. Therefore, we have mainly two subsystems: Front-end, which detects the landmarks of the map and correlates them with the poses, and the Back-end, which casts an optimization problem in order to estimate the pose and the location of landmarks.
    
    SLAM techniques can be distinguished to either Visual or LIDAR based Odometry (VO and LO) solutions, reflecting camera or LIDAR as the main sensor to be exploited:
    \begin{enumerate}
        \item To compute the local position and motion of a camera, VO algorithms must estimate the transformation that the camera undergoes between the current frame and a reference frame. The reference frame can be defined by the previous frame in the input sequence, some keyframe in the recent past, or a collection of frames from the recent past. In each case, the task is to estimate the transformation that takes information in the camera’s current frame into the frame of reference of the past frame(s). This task can be seen as an optimisation problem where the cost is given by the residual between the information measured in the current frame and corresponding information derived and reprojected from the reference frames. The vast majority of VO algorithms use feature-based approaches (e.g. \cite{Klein2007}). The image is decomposed to a sparse set of interesting points, where for each interest point location the local appearance of the image is described by a feature vector that is invariant to camera transformations. The feature vectors are associated between the input frame and reference to form a set of geometric constraints from which we can derive the camera motion and scene structure. In this case, the cost function is formulated by the difference between the measured reprojection location of these interest points between frames, referred to as the reprojection error. However, some known limitations of feature-based methods include i) extraction of interesting points and feature vectors may be expensive (using well-known algorithms like SHIFT or SURF), ii) they are prone to errors in areas where there is a low number of interesting points, etc. On the contrary, dense or direct VO approaches \cite{Steinbrucker2011, Whelan2013} focus on minimizing the (geometric) reprojection error, aiming to directly minimise the photometric error between pixels in the optimization problem. State-of-the-art VO algorithms include Direct Sparse Odometry (DSO) \cite{Engel2018}, ORB-SLAM \cite{MurArtal2015} and ORB-SLAM2 \cite{MurArtal2017}.
        \item LIDAR sensor provides dense 3D point clouds of vehicle's surroundings. The goal of LO is to estimate the pose of the vehicle by accumulating the transformation between consecutive frames of 3D point clouds. The existing LO solutions can be divided into two groups: point-wise and feature-wise methods. Point-wise methods estimate the relative transformation directly using the raw 3D points while feature-wise methods try to utilize more sophisticated characteristics of the point cloud such as the edge and planar feature points. The most well-known pointwise LO method is the iterative close point (ICP) \cite{Besl1992}. ICP operates at a point-wise level and directly matches two frames of the point cloud by finding the correspondences. One of the major drawbacks of the ICP is that when the frames include large quantities of points, ICP may suffer from a high computational load arising from the point cloud registration. Many variants of ICP have been proposed to improve its efficiency and accuracy, such as the Trimmed ICP \cite{makihara2002object} and Normal ICP \cite{Serafin2015}. To avoid the high computational load resulting from using the entire set of raw points, the feature-based LO methods extract a set of representative features from the raw points. The fast point feature histogram (FPFH) was proposed in \cite{Rusu2009} to extract and describe important features. The FPFH enables the exploration of the local geometry and the transformation is optimized by matching the one-by-one FPFH-based correspondence. Another well-known feature-based LO method is LOAM \cite{Zhang2014}. Theoretically, LOAM integrates the properties of both the point-wise and the feature-wise methods. On the one hand, to decrease the computational load of typical ICP, LOAM proposed to extract two kinds of feature points, the edge and planar, respectively. The extraction of the feature is simply based on the smoothness of a small region near a given feature point. Different from the FPFH which provides multiple categories of features based on its descriptors, LOAM involves only two feature groups. Another popular variant of LOAM, is Lego-LOAM \cite{Shan2018}. 
    \end{enumerate}
    
    \subsubsection{Path planning} Connected Advanced Driver Assistance Systems (ADAS) help to reduce road fatalities and have received considerable attention in the research and industrial societies \cite{7416035}. Recently, there is a shift of focus from individual drive-assist technologies like power steering, anti-lock braking systems (ABS), electronic stability control (ESC), adaptive cruise control (ACC) to features with a higher level of autonomy like collision avoidance, crash mitigation, autonomous drive and platooning. More importantly, grouping vehicles into platoons \cite{halder2020distributed,8731978} has received considerable interest, since it seems to be a promising strategy for efficient traffic management and road transportation, offering several benefits in highway and urban driving scenarios related to road safety, highway utility and fuel economy.  

    To maintain the cooperative motion of vehicles in a platoon, the vehicles exchange their information with the neighbours using V2V and V2I \cite{hobert2015enhancements}. 
    The advances in V2X communication technology \cite{hobert2015enhancements,8731978} enable
    multiple automated vehicles to communicate with one another,
    exchanging sensor data, vehicle control parameters and visually detected objects facilitating the so-called 4D cooperative awareness (e.g., identification/detection of occluded pedestrian, cyclists or vehicles). 
    
    Several works have been proposed for tackling the problems of cooperative path planning. Many of them focus on providing spacing policies schemes using both centralized and decentralized model predictive controllers. Though very few take into account the effect of network delays, which are inevitable and can deteriorate significantly the performance of distributed controllers. 
    
    The authors in \cite{8965227}, presented a unified approach to cooperative path-planning using Nonlinear Model Predictive Control with soft constraints at the planning layer. The framework additionally accounts for the planned trajectories of other cooperating vehicles ensuring collision avoidance requirements.  Similarly, a multi-vehicle cooperative control system is proposed in \cite{8410769,7331006} with a decentralized control structure, allowing each automated vehicle to conduct path planning and motion control separately. The authors in \cite{halder2020distributed}  present a robust decentralised state-feedback controller in the discrete-time domain for vehicle platoons, considering identical vehicle dynamics with undirected topologies. An extensive study of their performance under random packet drop scenarios is also provided, highlighting their robustness in such conditions. The authors in \cite{8710544} have extended decentralized MPC schemes to incorporate also the predicted trajectories of human driving vehicles. Such solutions are expected to enable the co-existence of vehicles supporting various levels of autonomy, ranging from L0 (manual operation) to L5 (fully autonomous operation) \cite{taeihagh2019governing}. 
    
    Additionally to the cooperative path planning mechanisms, spacing policies and controllers have also received increased interest towards ensuring collision avoidance by regulating the speeds of the vehicles forming a platoon. Two different types of spacing policies can be found in the literature, i.e., the constant-spacing policy \cite{7636964} and the constant-time-headway spacing policy (e.g., focusing on maintaining a time gap between vehicles in a platoon resulting in spaces that increase with velocity) \cite{7938659}. In both categories, most works, use a one direction control strategy. At this point, it should be mentioned that in a one-directional strategy the vehicle controller processes the measurements which are received from leading vehicles. Similarly, a bidirectional platoon control scheme takes into consideration the state of vehicles in front and behind (see \cite{6482670}). In most of the cooperative platooning approaches, the vehicle platoons are formulated as double-integrator systems that deploy decentralised bidirectional control strategies similar to mass–spring–damper systems. This model is widely deployed since it is capable of characterising the interaction of the vehicles with uncertain environments and thus is more efficient in stabilising the vehicle platoon system in the presence of modelling errors and measurement noise. Though, it should be noted that the effect of network delays on the performance of such systems, has not been extensively studied, despite the fact that time delays, including sensor detective delay, braking delay and fuel delay not only seems to be inevitable but also is expected to deteriorate significantly the performance of the distributed controllers.

\begin{table*}
	\caption{Cooperative Localization methods}
    \begin{center}
    	\centering
        \resizebox{\textwidth}{!}{
        \begin{tabular}{||c|c|c|c|c|c|c|} 
    		\hline
    		Fusion algorithm(s) & Survey & Centralized solution & Distributed solution & Benefits & Limitations & Reference\\ [0.5ex] 
    		\hline \hline 
    	     LS, GD and SPAWN & - & - & \checkmark & Two state-of-the art algorithms & 
    	     Large number of iterations and information exchange are required to reach good solution
    	     & Wymeersch et al. \cite{Wymeersch2009} \\
    	    \hline
    	     Particle filter based belief propagation & - & - & \checkmark & Distributed tracking of mobile nodes and non cooperative objects  & Nodes have to reach consensus on objects' position &  Meyer et al. \cite{Meyer2016} \\
    	    \hline
    	    EKF & - & - & \checkmark & Overall location estimation in harsh conditions and realistic network simulation  & Lacks evaluation for the individual vehicle  &  Elazab et al. \cite{Elazab2017} \\
    	    \hline
    	    Cubature KF and Huber M-estimation & - & - & \checkmark & Robust location estimation in the presence of measurement outliers  & Not considering the impact of dynamic VANET's topology  &  Liu et al. \cite{Liu2017} \\
    	    \hline
    	    - & \checkmark & - & - & Complete survey about the different fusion algorithms and technologies for CL & - &  Buehrer et al. \cite{Buehrer2018} \\
    	    \hline
    	     - & \checkmark & - & - & Complete survey about the different fusion algorithms and technologies for CL, including SLAM methods & - &   Kuutti et al. \cite{Kuutti2018} \\
    	     \hline
    	    Geometric algorithms & \checkmark & - & \checkmark & Linear and distributed approach based on sophisticated selection of neighbors & Developed mainly for static scenarios &  Safavi et al. \cite{Safavi2018} \\
    	    \hline
    	    Gaussian message passing and KF  & - &  \checkmark & \checkmark & Distributed CL method relying on the cooperatively detection of features & Vehicles have to reach consensus on features' position &  Soatti et al. \cite{Soatti2018} \\
    	    \hline
    	    - & \checkmark & - & \checkmark & Detailed book about the current and potential status of CL methods  & - &  Gao et al. \cite{Gao2019} \\
    	     \hline
    	    Particle filter based belief propagation & - & - & \checkmark & Distributed data association approach  & Vehicles have to reach consensus on features' position  &  Brambilla et al. \cite{Brambilla2020} \\
    	    \hline
    	    Graph Laplacian processing & - & \checkmark & - & Fusion of three measurement modalities via linear LS  & No motion model is concerned &  Piperigkos et al. \cite{Piperigkos2020} \\
    	    \hline
    	    Graph Laplacian processing & - & \checkmark & \checkmark & Fusion of three measurement modalities via linear LS  & No motion model is concerned  &  Piperigkos et al. \cite{Piperigkos2020b} \\
    	    \hline
    	    KF and ML & - & - & \checkmark & Effective and simple implementation of cooperative awareness & Measurement model is rather abstract, not discussing in detail how it can be formulated &  Yang et al. \cite{Yang2020} \\
    	    \hline
    	    Bayesian approach & - & - & \checkmark & Accurate location estimation in harsh conditions  & Only ego vehicle location is assessed &  Zhao et al. \cite{Zhao2020} \\
    	    \hline
    	\end{tabular}
    	}
    \end{center}
\end{table*}

\begin{table*}
	\caption{SLAM methods based on VO and LO solutions}
    \begin{center}
    	\centering
    	\begin{tabularx}{\textwidth}{||c|X|X|X|X|} 
    		\hline
    		Camera & LIDAR & Benefits & Limitations & Reference\\ [0.5ex] 
    		\hline \hline
    	    - & \checkmark & Fundamental work & High computational load & ICP \cite{Besl1992}\\
    	    \hline
    	    - & \checkmark & Variant of ICP & Improves the computational complexity of ICP & TICP \cite{makihara2002object}\\
    	    \hline
    		\checkmark & - & Fundamental feature based approach & Challenging the extraction of feature points & Klein et al. \cite{Klein2007}\\ 
    		\hline
    	    - & \checkmark & Exploits a set of representative features from raw point cloud & Lacks evaluation in different weather and lighting conditions & FPFH \cite{Rusu2009}\\
    	    \hline 
    		\checkmark & - & Directly minimize the photometric error between pixels & Sensitive to image noise & Steinbrucker et al. \cite{Steinbrucker2011}\\
    		\hline 
    		\checkmark & - & Directly minimize the photometric error between pixels & Sensitive to image noise & Whelan et al. \cite{Whelan2013} \\ 
    		\hline 
    	    - & \checkmark & State-of-the-art LO solution & Lacks evaluation in different weather and lighting conditions & LOAM \cite{Zhang2014}\\
    	    \hline 
    		\checkmark & - & State-of-the-art VO solution & Lacks evaluation in different weather and lighting conditions & ORB-SLAM \cite{MurArtal2015} \\
    		\hline 
    		- & \checkmark & Variant of ICP & Improves the computational complexity of ICP & NICP \cite{Serafin2015}\\
    	    \hline
    	    \checkmark & - & State-of-the-art VO solution & Lacks evaluation in different weather and lighting conditions & ORB-SLAM2 \cite{MurArtal2017}\\
    	    \hline
    		\checkmark & - & State-of-the-art VO solution & Lacks evaluation in different weather and lighting conditions & DSO \cite{Engel2018} \\
    		\hline
    	    - & \checkmark & State-of-the-art LO solution & Lacks evaluation in different weather and lighting conditions & LeGO-LOAM\cite{Shan2018}\\
    	    \hline
    	\end{tabularx}
    \end{center}
\end{table*}

\begin{table*}
	\caption{Path planning methods}
    \begin{center}
    	\centering
    	\begin{tabularx}{\textwidth}{||c|X|X|X||} 
    		\hline
    		Cooperative path planning & Spacing controller mechanism & Year & Reference\\ [0.5ex] 
    		\hline \hline
    	    - & \checkmark & 2013 & Ghasemi et al. \cite{6482670}\\
    	    \hline
    	    \checkmark & - & 2015 & Kuriki et al. \cite{7331006}\\
    	    \hline
    	     - & \checkmark & 2017 & Liu et al. \cite{7636964}\\
    	    \hline
    	    - & \checkmark & 2017 & Liu et al. \cite{7938659}\\
    	    \hline
    	    \checkmark & - & 2018 & Viana et al. \cite{8710544}\\
    	    \hline
    	    \checkmark & - & 2019 & Viana et al. \cite{8965227}\\
    	    \hline
    	    \checkmark & - & 2019 & Huang et al. \cite{8410769}\\
    	    \hline
    	    \checkmark & - & 2019 & Taeihagh et al. \cite{taeihagh2019governing}\\
    	    \hline
    	\end{tabularx}
    \end{center}
\end{table*}

\subsection{Human centric perception}
Human, as a part of a CPSoS, plays an important role to the functionality of the system. The humans' role in such complicated systems (e.g., CPSoS) is vital since they react and collaborate with the machines, providing them with useful feedback and affecting the way that these systems work. Humans can provide valuable input both in an active (on purpose) or in a passive (without consideration) way. For example, an input such as a gesture, or voice can be used as an order or command to control the operation of a system via a Human-Machine Interface (HMI). On the other hand, pose estimation or biometrics, like heart rate, could be taken into account by a decision component, resulting in a corresponding change of the system's functionality for security reasons (e.g., when a user's fatigue has been detected).

The following sections present some human-related inputs (e.g., behaviour, characteristics, etc.,) that can be beneficially used in CPSoSs, according to the literature.

\begin{itemize}
\item \textbf{Biometrics and Biometric Recognition}. The most well-known and most frequently used biometrics, related to humans, are face, fingerprint, iris, EEG, EGG, respiratory, and heart rate. Some of them are unique for each person so they can be used for human identification while others can be used for monitoring the humans' state or their special cognitive situation of a specific time period. The use of biometrics covers a large variety of tasks and applications in CPSoSs. 

Regarding the face of a human as a biometric, the related tasks can be face detection \cite{6664945,ISERN2020303,7524019,galbally2019study},
 face alignment \cite{9238315,9268760},
 face recognition \cite{9253224},
 face tracking \cite{7328312,7040663}, 
 face classification/verification \cite{10.1007/978-3-319-43958-7_62}, and
 face landmarks extraction \cite{8094282,8871109,8633862}. Fingerprint \cite{7977654,8681394,8952777}, palmprint \cite{1699574} and iris/gaze \cite{7053946,6826500} are mainly used for user's identification tasks due to their uniqueness for each person. EEG \cite{8399734,8759031,7886281}, EGG, respiratory \cite{9123889,9428504,9169861}, heart rate \cite{6526579,8538621,8754015} are used for the user's state monitoring. Besides the fact that they can provide valuable information, their usage in real applications is difficult to be applied due to the special wearable devices that it is required for the capturing. 
 
The choice of which specific biometric will be utilized depends on the use case scenario, the availability and feasibility of using a sensor (e.g., it will be placed in a stationary location or it has to be constantly wearable during the operation), the special power consumption needs of each sensor, the accuracy and the latency.

One other important issue, which needs to be taken under serious consideration before the use of a biometric in real systems, is the privacy and security of these sensitive data since they must be protected via encoding in order to be anonymously stored or used.

\item \textbf{Person Identification}. Person identification is a common image retrieval problem, where the objective of this task is the recognition of a specific person's identity by usually using only a single image, captured by a camera.

Generally, the person identification task is a more complicated and challenging problem in comparison with the identification using only the face, since face identification is applied in a more controlled environment (e.g., use of a smaller captured frame, the user has to remove glasses, hat and other accessories to be identified). On the other hand, person identification has to deal with more complex issues like the different points of view, light and weather conditions, different resolution of the camera, types of clothes and a large variety of background contexts.

Person identification has shown great usability in applications related to CPSoSs, mostly for security purposes. Its utility has been marked specifically when it is applied ``in the wild" and in uncontrolled environments where other biometrics are not feasible to be used due to technical constraints. Nowadays approaches usually use deep networks to perform reliable and accurate results.
	
Authors in \cite{9195852} proposed an additive distance constraint approach with similar labels loss to learn highly discriminative features for person re-identification.  In \cite{8976262}, the authors proposed a deep model (PurifyNet) to address the issue of the person re-identification task with label noise which has limited annotated samples for each identity. In \cite{8658110}, an unsupervised re-identification deep learning approach was used that is capable of incrementally discovering discriminative information from automatically generated person tracklet data. 

\item \textbf{Human Pose Estimation and Action Recognition}.  
	
Human pose estimation and recognition of the human's action have been proved as particularly valuable tasks in nowadays video-captured applications related to CPSoSs. They can be utilized in a variety of fields such as ergonomics assessment, safe training of new operators, fatigue and drowsiness detection of the user, human-machine interactions, prediction of operator's next action for avoiding accidents through changing the operation of a machine, dangerous moving monitoring in insecure workspace areas.

A restriction that can negatively affect and obstruct the quality of the results of these tasks is the limited coverage area of the camera. Nevertheless, this limitation can be overcome using new types of sensors and tools like Inertial Measurement
Units (IMUs), whole-body tracking system (e.g. SmartsuitPro, Xsens)  etc., \cite{7473872}.
	
In \cite{doi:10.1177/1687814019897228}, an approach is presented that exploits visual cues from human pose to solve industrial scenarios for safety applications in CPSs. In \cite{8453782}, three modalities (i.e., 3D skeletons, body part images and motion history image) are integrated into a hybrid deep learning architecture for human action recognition.
The authors in \cite{8441171} proposed a skeleton-based approach utilizing Spatio-temporal information and CNNs for the classification of human
activities. The authors in \cite{9211910} presented an indoor monitoring reconfigurable CPS that uses embedded local nodes (Nvidia Jetson TX2), proposing learning architectures to address Human Action Recognition.

\item \textbf{Hand Gesture Recognition}. 

Hand gesture recognition tasks can be a very useful tool for interactions with machines or subsystems in CPSoSs \cite{HORVATH2017184}, and particularly in applications where the user is not allowed to have physical hand contact with a machine due to security reasons. This task mainly consists of three sequential steps which are hand detection, hand tracking, and finally gesture recognition. This means that hand gesture recognition can occur either by a single image (i.e., static gesture recognition) or by a sequence of images (i.e., dynamic gesture recognition). The first strategy looks more like a retrieval problem where the gesture of the image has to match with a known predefined gesture from a dataset of gestures. The second is a more complicated problem but it is more useful since can cover the requirements of a bigger variety of real problems \cite{s17081893}. 	

Gesture recognition is a very common task in Human-Computer Interaction. Nonetheless, the recognition of complex patterns demands accurate sensors and sufficient computational power \cite{7496423}. Additionally, we have to refer that visual computing plays an important role in CPSoSs, especially in these applications where the visual gesture recognition system relies on multi-sensor measurements \cite{7064655, 1699081}.

Authors in \cite{HORVATH2017184} presented a control interface for cyber-physical systems that interprets and executes commands in a human-robot shared workspace using a gesture recognition approach. The authors in \cite{Lou2016} tried to address the problem of personalized gesture recognition for cyber-physical environments, proposing an event-driven service-oriented framework. While in other gesture recognition applications, a body-worn setup was proposed, which supplements the omnipresent 3 DoF motion sensors with a set of ultrasound transceivers \cite{9274795}.

\item \textbf{Speech and speaker recognition}.

Speech recognition is a sub-category of a more generic research area related to the domain of Natural Language Processing (NLP). The main objective of speech recognition is to automatically translate the content of the entire speech (or the most significant part of it) into text or other recognizable forms from the computers. Assuming that the recording and processing of speech do not require a special sensor, but just a simple audio recorder, we can understand how easy to use this information is. Additionally, speech can be applied without any physical contact interaction, making it an ideal signal for HMI applications.

Speech recognition tasks can be utilized in the
smart input system \cite{9270545,7423314}, 
automatic transcription system \cite{6338511,7296399},
smart voice assistant \cite{9210344}, 
computer-assisted speech \cite{9034084},
rehabilitation \cite{8554657,8579178} and 
language teaching.  
	
Similar to the face recognition task that focuses on the recognition of an individual human using the facial information that is enclosed in a single image, the speaker recognition task tries to achieve the same goal using the vocal tone information of the subject. Speaker recognition is one of the most basic components for human identification, which has various applications in many CPSoSs. Additionally, fusion schemes can be used combining both speaker recognition and face recognition for more secure integrations \cite{2011.08612}.

A speaker recognition system consists of three separate parts, namely the speech acquisition module, the feature extraction and selection module, and finally the pattern matching and classification module. In CPSoSs, the implementation of an automatic speech recognition system relies on a voice user interface so that humans to interact with robots or other CPS components. Nevertheless, this type of interface can not replace the classical GUIs but it can intensify them by providing, in some cases, a more efficient way of interaction.

The authors in \cite{8430295} developed a technique to train a Neural Network (NN) on the extracted Mel-frequency Cepstral Coefficient (MFCC) features from audio samples to increase the recognition accuracy of the short utterance speaker recognition system. In \cite{9291611}, the authors tried to improve the robustness of speaker identification, using a Stacked Sparse Denoising Auto-encoder.
	

\end{itemize}

\begin{table*}
	\caption{Datasets for face recognition, detection and facial landmarks extraction tasks.}
\begin{center}
	\centering
	\begin{tabularx}{\textwidth}{||c|X|X|X||} 
		\hline
		DATASET & Short Description & Link of the Dataset & Paper Name\\ [0.5ex] 
		\hline\hline
		 Helen & Helen dataset consists of 2330 images (400x400 pixels) with labeled facial components which are manually annotated, containing contours near to eyes, eyebrows, nose, lips and jawline. & \url{http://www.ifp.illinois.edu/~vuongle2/helen/} & Interactive Facial Feature Localization \cite{10.1007/978-3-642-33712-3_49}\\ 
		\hline
		AFW & AFW (Annotated Faces in the Wild) is a face detection dataset consisting of 205 images with 468 faces. Each face image is labelled with at most 6 landmarks with visibility labels, as well as a bounding box. 
		& \url{https://www.ics.uci.edu/~xzhu/face/} & Face detection, pose estimation, and landmark localization in the wild \cite{6248014} \\
		\hline
		300W  & 300-W dataset consists of 300 Indoor and 300 Outdoor ``in the wild" images, covering a large variety of identity, expression, illumination conditions, pose, occlusion and face size. & \url{https://ibug.doc.ic.ac.uk/resources/300-W/} & 300 Faces in-the-Wild Challenge: The First Facial Landmark Localization Challenge \cite{6755925} \\
		\hline
		LFPW  & The Labeled Face Parts in the Wild (LFPW) consists of 1,432 faces from images which are downloaded from the web (e.g., google.com, flickr.com, and yahoo.com). & \url{https://neerajkumar.org/databases/lfpw/} & Localizing parts of faces using a consensus of exemplars \cite{5995602} \\
		\hline
		AFLW & The Annotated Facial Landmarks in the Wild (AFLW) consists of 25,000 faces that are annotated with up to 21 landmarks per image.  The images have been gathered from Flickr, covering a large variety of poses, expressions, ethnicities, ages, genders and environmental conditions.  & \url{https://www.tugraz.at/institute/icg/research/team-bischof/lrs/downloads/aflw/} & Annotated Facial Landmarks in the Wild: A large-scale, real-world database for facial landmark localization \cite{6130513}\\  
		\hline
		AFLW2000-3D & AFLW2000-3D dataset consists of 2,000 images that have been annotated using 68 points representing 3D facial landmarks. This dataset is usually used for evaluation of 3D facial landmark detection models. & \url{http://www.cbsr.ia.ac.cn/users/xiangyuzhu/projects/3DDFA/main.htm} & Face Alignment Across Large Poses: A 3D Solution \cite{zhu2016face} \\ 
		\hline
		300-VW & 300 Videos in the Wild (300-VW) is a dataset for evaluating facial landmark tracking algorithms in the wild. Each video of this dataset is almost 1 minute in duration (at 25-30 fps). Each frame of all videos has been annotated in the same way as the 300 W dataset.  & \url{https://ibug.doc.ic.ac.uk/resources/300-VW/}  &  Offline Deformable Face Tracking in Arbitrary Videos \cite{7406475} \\
		\hline
		COCO-WholeBody & This dataset is an extension of COCO dataset cavering a whole-body annotation (i.e., face, hand, feet) & \url{https://github.com/jin-s13/COCO-WholeBody} & Whole-Body Human Pose Estimation in the Wild \cite{10.1007/978-3-030-58545-7_12} \\
		\hline
		MALF  & MALF consists of 5,250 images with 11,931 faces in total. This dataset is the first face detection dataset that supports fine-gained evaluation.  & \url{http://www.cbsr.ia.ac.cn/faceevaluation/} & Fine-grained Evaluation on Face Detection in the Wild \cite{faceevaluation15} \\
		\hline
		FDDB & FDDB dataset consists  of 2,845 images with 5,171 annotated faces. & \url{http://vis-www.cs.umass.edu/fddb/index.html} & FDDB: A Benchmark for Face Detection in Unconstrained Settings \cite{fddbTech} \\
		\hline
	\end{tabularx}
\end{center}
\end{table*}

\begin{table*}
	\caption{Datasets of images with iris.}
	\begin{center}
		\begin{tabularx}{\textwidth}{||c|X|X|X||} 
			\hline
			DATASET & Short Description & Link of the Dataset & Paper Name \\ [0.5ex] 
			\hline\hline
			UBIRIS.v2 & The UBIRIS.v2 dataset consists of 11,102 images of iris that were captured from 261 subjects, with 10 images for each subject. The images were acquired using a variety of different conditions like distance, motion and different visible wavelengths. They have been also affected by real noise.  &  \url{http://iris.di.ubi.pt/ubiris2.html} & The UBIRIS.v2: A Database of Visible Wavelength Iris Images Captured On-the-Move and At-a-Distance \cite{4815254} \\ 
			\hline
			OpenEDS & Open Eye Dataset (OpenEDS) consists of images with eyes captured using a virtual-reality head display. This dataset was collected from 152 individual participants and is divided into four subsets. 
			& \url{https://research.fb.com/programs/openeds-challenge} &  OpenEDS: Open Eye Dataset \cite{1905.03702} \\
			\hline
		\end{tabularx}
	\end{center}
\end{table*}

\begin{table*}
	\caption{Datasets for pose estimation.}
	\begin{center}
		\begin{tabularx}{\textwidth}{||c|X|X|X||} 
			\hline
			DATASET & Short Description & Link of the Dataset & Paper Name \\ [0.5ex] 
			\hline\hline
			COCO  & The Microsoft Common Objects in Context (MS COCO) consists of 328,000 images. This dataset is a general-proposed large-scale object detection, segmentation, key-point detection, and captioning dataset containing also labeled human's poses.  & \url{https://cocodataset.org/} & Microsoft COCO: Common Objects in Context \cite{10.1007/978-3-319-10602-1_48} \\ 
			\hline
			MPII & The MPII Human Pose Dataset consist of 25,000 images of which 15,000 images are training samples, 3,000 images are validation samples and the rest 7,000 images are testing samples. The single-person poses are manually annotated with up to 16 body joints. The images are taken from YouTube videos covering 410 different human activities. & \url{http://human-pose.mpi-inf.mpg.de/}  & 2D Human Pose Estimation: New Benchmark and State of the Art Analysis \cite{10.1109/CVPR.2014.471} \\
			\hline
			DensePose & DensePose-COCO is a large-scale ground-truth dataset with image-to-surface correspondences which are manually annotated from 50,000 images of the COCO dataset and train DensePose-RCNN, to densely regress part-specific UV coordinates within every human region at multiple frames per second. & \url{http://densepose.org/} & DensePose: Dense Human Pose Estimation in the Wild \cite{8578860} \\
			\hline
			LSP & The Leeds Sports Pose (LSP) dataset consists of 2,000 images of sportspersons in total gathered from Flickr, 1,000 for training and 1,000 for testing. This dataset is used for human pose estimation and each image is annotated with 14 joint locations. & \url{https://dbcollection.readthedocs.io/en/latest/datasets/leeds\_sports\_pose\_extended.html} & Clustered Pose and Nonlinear Appearance Models for Human Pose Estimation \cite{BMVC.24.12} \\
			\hline
			JHMDB  & JHMDB is a recognition dataset that consists of 960 video sequences belonging to 21 actions. This dataset is a subset of the larger HMDB51 dataset which has been collected from digitized movies and YouTube videos. & \url{http://jhmdb.is.tue.mpg.de/} & Towards Understanding Action Recognition \cite{6751508} \\
			 \hline
			 Unite the People & Unite The People dataset is mainly used for 3D body estimation. The images come from an extended version of LSP dataset, as well as the single person-tagged people from the MPII Human Pose Dataset. The images are labeled with different types of annotations such as segmentation labels, pose or 3D representation. & \url{https://files.is.tuebingen.mpg.de/classner/up/} & Unite the People: Closing the Loop Between 3D and 2D Human Representations \cite{8099983} \\
			  \hline
		\end{tabularx}
	\end{center}
\end{table*}

\begin{table*}
	\caption{Datasets for hand and gesture recognition.}
	\begin{center}
		\begin{tabularx}{\textwidth}{||c|X|X|X||} 
			\hline
			DATASET & Short Description & Link of the Dataset & Paper Name\\ [0.5ex] 
			\hline\hline
			HandNet & The HandNet dataset contains the depth images of 10 participants' hands non-rigidly deforming in front of a RealSense RGB-D camera. The annotations were generated by a magnetic annotation technique. 6D pose is available for the center of the hand as well as the five fingertips (i.e. position and orientation of each). & \url{http://www.cs.technion.ac.il/~twerd/HandNet/} & Rule of thumb: Deep derotation for improved fingertip detection \cite{BMVC2015_33} \\ \hline
			EgoGesture & The EgoGesture dataset consists of 2,081 RGB-D videos, 24,161 gesture samples and 2,953,224 frames from 50 distinct subjects. & \url{http://www.nlpr.ia.ac.cn/iva/yfzhang/datasets/egogesture.html} & EgoGesture: A New Dataset and Benchmark for Egocentric Hand Gesture Recognition \cite{8299578} \\ \hline
			NVGesture & The NVGesture dataset consists of 1,532 dynamic gestures categorized into 25 classes. The dataset is separated into 1,050 samples for training and 482 for testing. The application in which it can be used is for touchless driver controlling. & \url{https://research.nvidia.com/publication/online-detection-and-classification-   dynamic-hand-gestures-recurrent  -3d-convolutional} & Online Detection and Classification of Dynamic Hand Gestures With Recurrent 3D Convolutional Neural Network \cite{7780825} \\ \hline
			IPN Hand & The IPN Hand is a dataset consisting of videos with sufficient size, variation, and real-world elements capable to be used by deep neural networks for training and evaluation. The application on which this dataset focuses is dynamic hand gesture recognition. & \url{https://github.com/GibranBenitez/IPN-hand} & Real-time Hand Gesture Detection and Classification Using Convolutional Neural Networks \cite{8756576} \\ \hline
			MLGESTURE  & MlGesture consists of more than 1300 hand gesture videos from 24 participants and features 9 different hand gesture symbols. The dataset has been recorded in a car with 5 different sensor types at two different viewpoints and it can be used for hand gesture recognition tasks.  & \url{https://iiw.kuleuven.be/onderzoek/eavise/mlgesture/home} & Low-latency hand gesture recognition with a low resolution thermal imager \cite{9150613} \\ \hline
\end{tabularx}
\end{center}
\end{table*}

\begin{table*}
	\caption{Datasets of action recognition.}
	\begin{center}
		\begin{tabularx}{\textwidth}{||c|X|X|X||} 
			\hline
			DATASET & Short Description & Link of the Dataset & Paper Name \\ [0.5ex] 
			\hline\hline
			UCF101 & This dataset consists of 13,320 video clips ($\sim$ 27 hours) from Youtube, classified into 101 categories and into 5 types (i.e., Body motion, Human-human interactions, Human-object interactions, Playing musical instruments and Sports). & \url{https://www.crcv.ucf.edu/data/UCF101.php} & UCF101: A Dataset of 101 Human Actions Classes From Videos in The Wild \cite{1212.0402} \\ 
			\hline
			Kinetics & It is a high-quality dataset of videos used for human action recognition. The dataset consists of around 500,000 labeled video clips of 10 seconds covering 600 human action classes with at least 600 video clips for each action class.  & \url{https://deepmind.com/research/open-source/kinetics} & The Kinetics Human Action Video Dataset \cite{1705.06950} \\ 
			\hline						
			HMDB51 & The HMDB51 is a dataset consisting of 6,849 video clips from 51 action categories (such as “jump”, “kiss” and “laugh”). Each category containing at least 101 clips. & \url{https://serre-lab.clps.brown.edu/resource/hmdb-a-large-human-motion-database/} & HMDB: A large video database for human motion recognition \cite{6126543} \\ 
			\hline
			ActivityNet & The ActivityNet contains 200 different types of activities and a total of 849 hours of videos collected from YouTube. It is one of the largest datasets regarding the number of activity categories and a number of videos. & \url{http://activity-net.org/} & ActivityNet: A Large-Scale Video Benchmark for Human Activity Understanding \cite{7298698} \\ 
			\hline
				NTU RGB+D & NTU RGB+D consists of 56,880 video clips of 60 action classes collected from 40 subjects. The actions can be generally divided into three categories: 40 daily actions (e.g., drinking, eating, reading), nine health-related actions (e.g., sneezing, staggering, falling down), and 11 mutual actions (e.g., punching, kicking, hugging).  & \url{http://rose1.ntu.edu.sg/datasets/actionrecognition.asp} & NTU RGB+D: A Large Scale Dataset for 3D Human Activity Analysis \cite{1604.02808} \\ 
				\hline
				KTH &  The KTH dataset contains six actions: walk, jog, run, box, hand-wave, and hand clap by 25 different individuals, in different environments outdoor (s1), outdoor with scale variation (s2), outdoor with different clothes (s3), and indoor (s4).  & \url{https://www.csc.kth.se/cvap/actions/} & Recognizing Human Actions: A Local SVM Approach \cite{1334462} \\ 
				\hline
				Composable activities dataset & This dataset consists of 693 annotated videos of activities in 16 classes performed by 14 individuals. & \url{https://ialillo.sitios.ing.uc.cl/ActionsCVPR2014/} & Discriminative Hierarchical Modeling of Spatio-Temporally Composable Human Activities \cite{6909504} \\ 
				\hline
				HACS  &  HACS dataset contains 504K videos (shorted than 4 minutes) collected from YouTube, categorized in 200 action classes. It is used human action recognition.  & \url{http://hacs.csail.mit.edu/} & HACS: Human Action Clips and Segments Dataset for Recognition and Temporal Localization \cite{zhao2019hacs} \\ 
				\hline
		\end{tabularx}
	\end{center}
\end{table*}

\begin{table*}
	\caption{Datasets for speech recognition.}
	\begin{center}
		\begin{tabularx}{\textwidth}{||c|X|X|X||} 
			\hline
			DATASET & Short Description & Link of the Dataset & Paper Name \\ [0.5ex] 
			\hline\hline
			LibriSpeech & This dataset consist of approximately 1,000 hours of audiobooks. &  \url{http://www.openslr.org/12} &  Librispeech: An ASR corpus based on public domain audio books \cite{7178964} \\ 
			\hline
			Speech Commands & Speech Commands consists of 65,000 of 30 short words $\sim$ one second long. It is a collection of spoken words by thousands of different people, designed for the training and evaluation of keyword spotting systems. &  \url{https://ai.googleblog.com/2017/08/launching-speech-commands-dataset.html} & Speech Commands: A Dataset for Limited-Vocabulary Speech
			Recognition \cite{1804.03209} \\ 
			\hline
			MuST-C & MuST-C currently represents the largest publicly available multilingual corpus for speech translation from English into several languages. It covers eight languages. It consists of hundred hours of audio recordings from English TED Talks. &  \url{https://ict.fbk.eu/must-c/} & MuST-C: A multilingual corpus for end-to-end speech translation \cite{CATTONI2021101155} \\ 
			 \hline
			 Common Voice & Common Voice is a dataset of 9,283 recorded hours that consists of audio files and corresponding text files including demographic metadata like age, sex, and accent.  &  \url{https://commonvoice.mozilla.org/en/datasets} & Common Voice: A Massively-Multilingual Speech Corpus \cite{1912.06670} \\ 
			  \hline
			 Libri-Light & Libri-Light is a collection of over 60K hours of spoken English suitable for training speech recognition systems under limited or no supervision.  &  \url{https://github.com/facebookresearch/libri-light} & Libri-Light: A Benchmark for ASR with Limited or No Supervision \cite{kahn2020libri} \\ 
			 \hline
			 THCHS-30 & THCHS-30 is a free Chinese speech database that can be used for speech recognition systems. &  \url{http://166.111.134.19:7777/data/thchs30/README.html} & THCHS-30 : A Free Chinese Speech Corpus \cite{1512.01882} \\ 
			 \hline
			 VOICES & This dataset consists of speech recorded by far-field microphones in noisy room conditions for using in speech and signal processing approaches.
			  &  \url{https://registry.opendata.aws/lab41-sri-voices/} & Voices Obscured in Complex Environmental Settings (VOICES) corpus \cite{1804.05053} \\ 
			 \hline    	
			 LibriCSS & LibriCSS is a real recorded dataset that simulates conversations where are captured by far-field microphones.  &  \url{https://github.com/chenzhuo1011/libri_css} & Continuous speech separation: dataset and analysis \cite{2001.11482} \\ 
			\hline    		
		    SPEECH-COCO & SPEECH-COCO contains 616,767 audios generated using text-to-speech (TTS) synthesis. The audio files are paired with images. &  \url{https://zenodo.org/record/4282267} & PEECH-COCO: 600k Visually Grounded Spoken Captions Aligned to MSCOCO Data Set \cite{havard2017speech} \\ 
		    \hline    			 		
		\end{tabularx}
	\end{center}
\end{table*}

\section{Behavioral layer}

In each CPSoS, the knowledge, senses and expertise of humans constitute important informative values that can be taken into account for the insurance of its operational excellence. However, a major concern, which was needed to be addressed in an early age of CPSoS's evolution, is in which way these human abstract features can be accessible and understandable by the CPSoSs. 

A way to enter the human, as a separate component into a CPSoS, is by introducing an anthropocentric mechanism which is known in the literature as the human-in-the-loop approach \cite{Gaham2015}\cite{hadorn2016towards}. This mechanism allows a direct way that humans can continuously interact with the CPSoSs' control loops in both directions of the system (i.e., taking and giving inputs).

Even though common CPSoSs are human-centred systems, in which the human constitutes an essential part of the system, unfortunately, in many real cases, these systems still consider the human as an external and unpredictable element, without taking its importance into deeper consideration. The main vision of the researchers and engineers is to create a human-machine symbiosis, integrating humans as holistic beings within CPSoSs. In this way, CPSoSs have to support a more tight bond with the human element, through human-in-the-loop controls taking into account human's features like intents, psychological and cognitive states, emotions and actions that all of them can be deduced through sensors data and signal processing approaches.

Additionally, engineers, which design and develop new generations CPSoSs, have to understand and realize which are these specific features of the CPSoSs that make them be different from the traditional CPSs. One of these things is the human-in-the-loop mechanism that allows to CPSoS taking advantage of some special humans' characteristics, which make them superiors in comparison with the machines. The technological assessments are not mature yet to directly integrate these human-oriented characteristics into machines and robots, so the use of the human-in-the-loop component is essential to serve the initial goals of a CPSoS. These characteristics, as have been proposed by the literature \cite{7389271}, are presented below:

\begin{itemize}
	\item \textbf{Cognition}. Humans have a totally different way to observe a situation, understand a problem and make final decisions based on lacking data than computers do. Human's cognition is the combined result of knowledge, experience, inspiration and intuition where no nowadays machine can overcome or even approach in some way.
	\item \textbf{Predicatibility}. People are not programmed to perform the same task, in the same way, every time that they try. In some cases, this would be a problem, especially when they have to follow very specific instructions. This characteristic might make them less reliable than a simple computer that just follows precise orders. However, this unpredictable behaviour could be beneficial in a critical situation, which may suddenly appear and that has not been distinctly defined in the script of the instructions. The ability of humans to be easily adapted to unknown situations make them a perfect component to provide out-of-the-box solutions in hazardous circumstances.
	\item \textbf{Motivation}. Humans, from their nature, usually require incentives and become more productive when they assure them. Motivation can guide a human to perform more effort to a task than what is required. On the other hand, computers and machines follow a very specific pipeline of work and they are not able to change the way that they perform a task in order to enhance their productivity.
\end{itemize}

The human-in-the-loop applications can be separated into three main categories in respect to the type of input that human provides: 
\begin{enumerate}
	\item these applications that the human has a head role and directly control the functionality of the CPSoS as a user or operator 
	\item these applications where the system passively
	monitors humans (e.g., biometrics, pose e.g.,) and based on them it takes decisions for appropriate actions
	\item these which are a hybrid combination of the aforementioned two types
\end{enumerate}

\textbf{Direct human control of a CPSoS}. The applications of this category can be separated into two different sub-categories, related to the degree of freedom that the CPSoS has in order to make its own decision and takes unsupervised actions. 

In the first sub-category, operators manage a process that is close to an autonomous task. This means that the system has full control of its action however the user is responsible to adjust some parameters that may affect the functionality of the system when it is required for external reasons. An example that can describe a scenario like this, is when an operator sets new values, to specific parameters on a machine in the industry, for changing the operation of the assembly line (e.g., for a new product). 

In the second sub-category, the operator has even more power and plays a more active role in the process, by directly controlling some tasks, setting explicit commands for the operation of the machines or robots. An example of this scenario is when an expert operator has to take complete control of a robotic arm remotely, for repairing purposes.

\textbf{Human monitoring applications}. The applications of this category represented by systems that passively monitor humans' actions, behaviour and biometrics using the acquired data of sensors to make appropriate decisions or to display the information. Based on the type of reaction of the system, the applications can be separated into two types, namely the open-loop and closed-loop systems. 

Open-loop systems continuously monitor humans and just visualize (e.g., smart glasses) or send a report with appropriate results, which may be useful or interesting for the operator. The system does not take any further action in this case. The presented results can cover (i) the first level of information, (ii) the second level of information and (iii) KPIs. First level information includes measurements that usually are directly received by the sensors (e.g., heart rate, respiratory, blinking of eyes). After the appropriate process of the first level information, it can be produced the second level information (e.g. drowsiness, awareness, anxiety level) that corresponds to a higher contextual meaning. 


Closed-loop systems use the received information of the sensors and the processing results in order to take an action. For example in an automotive use case, if critical drowsiness of a driver is detected by the system then the car would take full control of the vehicle and additionally would appropriately inform the driver of his condition, as a supplementary task.


\textbf{Hybrid Systems} Hybrid systems take human-centric sensing information as feedback to perform an open or closed-loop action but additionally, at the same time, they also take into account the direct human inputs and preferences. An example in the manufacturing use case, the system monitors the operator's actions, while he collaborates with a robot, and provides appropriate guided instructions. However, the level of detail of the guided assistant can be modified by the personalized preferences of the user that can be related by the level of his experience. 
\\\\
Humans play an important role in CPSoSs. Their contribution can be summarized into three categories (i) for data acquisition, (ii) for inference related to their state and (iii) for the actuation of an action to complete a task of their own or to collaborate with other components of the system \cite{7029083}. More specifically:

\begin{enumerate}
	\item Data acquisition:
	\begin{itemize}
		\item Human as an informer. Humans provide the system with information through the wearable sensors that may carry or through other devices that monitor them.
		\item Human as a communicator to transfer condensed knowledge. Humans are excellent communicators. They have the special ability to easily understand complicated information, make conclusions and pass to the system with filtered, useful and deductive information.
	\end{itemize}
	\item State inference:
	\begin{itemize}
		\item Human as an insider component. Training algorithms and machine learning approaches can be used for the recognition of the human's state (e.g., cognitive, physical, emotional, phycological, etc.,) which may affect the good functionality of a CPSoS or put the safety of the user at risk. When a problematic state of a user is identified, the system can change the typical operation that runs, with the purpose to protect him, or just to inform him with an appropriate message or warning. 
		\item Human as a feedback component. Based on the state of the user, the system may provide suggestions or recommendations to them. The acceptance of these suggestions by the users can be further utilized by the system as useful feedback to relate user's preferences and corresponding user's states, providing more personalized solutions in future similar situations. 
	\end{itemize}
	\item Actuation:
	\begin{itemize}
		\item Human as actuators. The actions that a human, as a part of a CPSoS, can do, are: (i) to set the values of some parameters,  (ii) to execute specific tasks or (iii) to take the entire control of the system, if required.
	\end{itemize}	
\end{enumerate}



\textbf{Human-machine-interface (HMI)}  is referred to as the medium that is utilized for the direct communication between humans and machines. This interface is used to facilitate a physical interaction between humans and machines \cite{Ajoudani2018}. Typically, the classical HMI system makes up of some common hardware components, like a screen and keyboard, and software with specialized functionalities, performing in this way a graphical user interface (GUI). All sensors and wearable devices, which are connected with humans or other components of the CPSoS, are also part of an HMI. The HMI has very extensive usage in CPSoSs, allowing each part of the CPSoS to directly interact with a human and vice versa, creating a synergy loop between cyber-physical systems and humans. In the future, HMIs will be also able to have social cohesion between humans and machines.



The authors in \cite{6945523} suggested that the main representative of HMI tools in CPSoSs, which are mainly used for the communication between humans and machines, are automatic speech recognition, gesture recognition, and extended reality, which can be represented by augmented or virtual reality. In such implementations, a touch screen can be the place where humans and machines meet each other in order to visualize, manipulate and exchange their goals, models, and ideas. 
The authors in  \cite{7946695}, presented a framework that is capable of visually acquiring information from HMIs in order to detect and prevent human-in-the-loop errors in the control rooms of Nuclear Power plants.
The intelligent and adaptable CPSoSs expect the automation systems to be decentralized and support ”Plug-and-Produce” features.
In this way, the HMIs have to dynamically update and adapt display screens and support elements to facilitate the work of the operators like IO fields and buttons \cite{9212011}.
The authors in \cite{8875495}, proposed a graphical HMI mechanism for intelligent and connected in-vehicle systems in order to offer a better experience to automotive users. While in \cite{10.5555/3108905.3108906}, the authors presented a way of connecting an HMI with a software model of an embedded control system and thermodynamic models in a hybrid co-simulation.

\textbf{Augmented reality (AR)} is the technological tool that allows the enrichment of the physical world utilizing digital content and information superimposed on top of a perceived representation of the real environment Theoretically, this means that any of the senses that a human has would potentially benefit from this technology. Nevertheless, in practice, real applications, using augmented reality technology, usually render visual, audio or haptic virtual information to the physical space of the user.

Recently, modern AR devices (like goggles, smart glasses, head-mounted
displays (HMD) and tablets) have been tremendously evolved to provide great functionalities to their users. However, there are still open issues related to the problem of how comfortable and
productive can be their use. To overcome these limitations, authors in  \cite{8390759} proposed a mechanism, namely “accented visualization” that allows adapting additional data, presented by AR devices, according to the user’s current interest, attention and focus.

In \cite{7011609}, a paradigm of CPSs is presented showing how it can be implemented in the pre-construction industry by integrating it with the AR technology to make real-time decisions. Linking AR with CPS offers promising perspectives for process-centric guidance of CPS users, where manual tasks can be guided and monitored, enabling
a much better traceability \cite{8501479}. However, AR applications are still not widely used in industry, which may lay at the complex industrial
requirements (e.g., technical, environmental and
regulative) \cite{8699183}. 

\textbf{Prediction of operator’s intentions} is a task that can improve the effectiveness of collaboration between CPSoSs and humans. An accurate prediction can be very essential, especially in industrial scenarios where the resilience and safety of all CPSoSs components mostly depend on the mutual understanding between humans and CPSoSs. So, it seems necessary to design and develop reliable, robust and accurate human behaviour modelling techniques that are capable to predict the human's actions or behaviour. 

On the one hand, operators are mainly responsible for their own safety when they are in the same working environment with a cobot, performing collaborative tasks. However, on the other hand, CPSoSs must have smart components that are able to identify, understand and even predict operators' intentions with a primary goal to protect them from a serious injury, for instance. For this implementation, a continuous video capturing component can be used by the prediction system to detect, track, recognize human's gestures or postures and an artificial intelligence component to predict human's intention.
		
The system can anticipate when unexpected human operations have been detected or specific human activity patterns have been predicted \cite{8542683}. In the meanwhile, the cobot can perform other tasks \cite{GARCIA2019600}. In literature, a lot of different approaches have been presented to solve the problem of prediction operator's intentions, such as a framework for the prediction of human intentions from RGBD data \cite{8433892}. A sparse Bayesian learning-based human intention predictor to predict the future human desired position \cite{8859792}. A temporal CNN with a convolution operator for human trajectory prediction \cite{9309403}. A system that detects human intentions through a recursive Bayesian classifier, exploiting head, and hand tracking data \cite{8433892}. A human intention inference system that uses Expectation-Maximization algorithm with online model learning \cite{7797625}.


\textbf{Awareness}

Situation Awareness is used to describe the level of awareness that operators/drivers/users have of the situation in order to perform tasks successfully \cite{doi:10.1518/001872095779049543}. Based on the definition in \cite{vidulich1994situation}, situational awareness needs to include four specific requirements:
\begin{enumerate}
	\item to easily receive information from the environment.
	\item to integrate this information with relevant internal knowledge, creating a mental model of the current situation.
	\item to use this model to direct further perceptual exploration in a continual perceptual cycle.
	\item to anticipate future events.
\end{enumerate}

Taking these four requirements into account, situational awareness is defined as the continuous extraction of environmental information, the integration of this information with previous knowledge to form a coherent mental picture, and the use of that picture in directing further perception and anticipating future events. The system will be able to monitor and understand the user's state (e.g. fatigue, cognitive level, etc) in order to produce personalized alarms, warnings, information and suggestions to the users. A situational awareness application could also provide:
\begin{itemize}
	\item Information streams regarding the task underway improving focus
	\item Personalized reminders regarding other parallel or scheduled tasks significantly improving response time
	\item Notifications and visual aids regarding imminent dangers or accident-related factors
	\item Environmental values and real-time measurements of sensors 
	\item KPIs visualizing the effectiveness of the CPSoS functionality
\end{itemize}

Situation awareness is important in cases where a user must intervene in operations and cooperations of highly automated systems in order to correct failed autonomous decisions in CPSoSs \cite{9108728}. It is also an effective method to keep the mechanical parts of a system as well as the operators secure and safe, so that it can be classified into two groups, human or computer awareness \cite{8448886}. Moreover, situational awareness for security reasons is very important, since it can be used to inform the user about an attack that takes place in real-time  \cite{8436591}.

Nevertheless, CPSoSs have also to overcome some challenges in respect of
human-in-the-loop component, which are:

\begin{itemize}
\item \textbf{Processing in real-time.} The complexity of CPSoSs, consisting of a variety of different components, leads to the production of a huge amount of data instantly. The processing of all these data and the real-time decisions making are challenging tasks, especially when the human-in-the-loop component is apparent since human's safety is the most important issue and the processes related to it would be handle with distinct sensitivity. 

The processing of data in batches could be a solution to this challenge. However, this approach is not reliable in critical situations of CPSoSs that may appear, where vital and accurate decisions have to be made quickly to protect human life and security. So in other words, the real-time data processing framework requires the ability of the system to handle large amounts of data with very low latency and in relative high performance.

\item \textbf{Online streaming of data.} CPSoSs is a system of systems that are connected to each other, collaborating and transferring also in real-time useful information and data. The requirement of real-time processing results in the need for the processing of these data in an online streaming mode. 

The challenge, in the case of online streaming, is due to the fact that data is transferred in an ordered sequence of instances that usually can be accessed once or a small number of times due to limited computing and storage capabilities. 

The tremendous growth of data demands switching from traditional data processing solutions to systems, which can process a continuous stream of real-time data.



\item \textbf{High-dimensional data.} High-dimensional data are referred to those data whose dimensions are more complex than the ordinary data. The presence of high-dimensional data is becoming a very common issue in many real-world applications of CPSoSs.  The processing of high-dimensional data that has been acquired by different sensors and devices presents a fundamental challenge leading to the need for more sophisticated methods to be developed.

High-frequency data are referred to as those data that usually appear as time series and the update of their values happens very fast (i.e., new observations take place every milliseconds-second).  The appropriate handle of high-frequency data is essential for contemporary CPSoSs. 
 
Processing of these data introduces new challenges to decision-making tasks, especially when a human takes part in the CPSoS as a human-in-the-loop component.

\item \textbf{Unsupervised learning in data of CPSoSs.} Unsupervised learning is a type of learning that tries to autonomously discover hidden patterns in untagged data. This is a very useful method to be applied in real-time applications where the observed data has a large variety in comparison with those of a restricted dataset. However, at the same time is a very challenging task when it is applied in CPSoSs that require accurate and precise results and usually, there are no ``ground truth" data for the evaluation of the method's accuracy \cite{Ma2018}.
\end{itemize}

\section{Conclusion}

This survey aimed to provide a comprehensive review on current best practices in connected cyberphysical systems adopting a dual architecture approach with a perception and a behavioral layer. Cooperative path planning is also discussed in the context of autonomous vehicles. Several works have been proposed for tackling the problems of cooperative path planning. Many of them focus on providing spacing policies schemes using both centralized and decentralized model predictive controllers. Though very few take into account the effect of network delays, which are inevitable and can deteriorate significantly the performance of distributed controllers. The individual nodes solve a simultaneous localization and mapping (SLAM)  problem refering to the problem of mapping an environment using measurements from sensors (e.g. Camera or LIDAR) on-board the vehicle or robot, whilst at the same time estimating the position of that sensor relative to the map. Object detection from either 2D imagery and 3D LIDAR point clouds provide range measurements through the application of deep neural networks.

In the presented setup, CPSoS depend on humans since humans are part of the CPSoS functionality and services, the interact with the CPSs and contribute to the CPSoS behavior. Human, as a part of a CPSoS, plays an important role to the functionality of the system. The humans' role in such complicated systems (e.g., CPSoS) is vital since they react and collaborate with the machines, providing them with useful feedback and affecting the way that these systems work. Humans can provide valuable input both in an active (on purpose) or in a passive (without consideration) way.  Operators and managers play a key role in the operation of CPSoS and take many important decisions while in several cases human CPS users are key player in the CPSoS main role thus forming Cyber Physical Human Systems. Human-in-the-loop approach refers to an anthropocentric mechanism that allows a direct way that humans can continuously interact with the CPSoSs' control loops in both directions of the system (i.e., taking and giving inputs).

\bibliography{survey_v1}
\bibliographystyle{IEEEtran}

\end{document}